\documentclass{aa}
\usepackage{graphics}
\usepackage{epsfig}

\def\php{\phantom{.}}

\begin{document}
\thesaurus{3(11.01.2; 11.09.1 B1144+352, B1144+353; 11.10.1; 13.18.1; 13.25.2)}  
\title{The Mpc-scale radio source associated with the GPS galaxy B1144+352}

\author{A.P. Schoenmakers\inst{1,2}
\and A.G. de Bruyn\inst{3,4}
\and H.J.A. R\"{o}ttgering\inst{2}
\and H. van der Laan\inst{1}} 
\institute{Astronomical Institute, Utrecht University, P.O. Box 80\,000, 3508~TA Utrecht, The Netherlands 
\and {Sterrewacht Leiden, Leiden University, P.O. Box 9513, 2300~RA Leiden, The Netherlands}
\and {NFRA, P.O. Box 2, 7990~AA Dwingeloo, The Netherlands}
\and {Kapteyn Astronomical Institute, University of Groningen, P.O. Box 800, 9700~AV Groningen, The Netherlands}}
\offprints{A.P. Schoenmakers (Schoenma@phys.uu.nl)}
\date{Received ; accepted}
\titlerunning{GPS galaxy B1144+352}
\authorrunning{A.P. Schoenmakers et al.}

\maketitle

\begin{abstract}

We present the results of new observations of the enigmatic radio source B1144+352 with the WSRT at 1.4 GHz.
This source is hosted by an $m_r = 14.3 \pm 0.1$ galaxy at a redshift of $z=0.063 \pm 0.002$ and is one of the lowest redshift Gigahertz Peaked Spectrum (GPS) sources known. 
It has been known to show radio structure on pc-scale in the radio core and on $20-60$ kpc-scale in two jet-like radio structures. The WENSS and NVSS surveys have now revealed faint extended radio structures on an even much larger scale. We have investigated these large-scale radio components with new 1.4-GHz WSRT observations.
Our radio data indicate that the eastern radio structure has a leading hotspot and we conclude that this structure is a radio lobe originating in the galaxy hosting the GPS source. The western radio structure contains two separate radio sources which are superposed on the sky. The first is a low-power radio source, hosted by a $m_R = 15.3 \pm 0.5$ galaxy at a similar redshift ($z=0.065\pm0.001$) to the GPS host galaxy; the second is an extended radio lobe, which we believe is associated with the GPS host galaxy and which contains an elongated tail.
The total projected linear size of the extended radio structure associated with B1144+352 is $\sim\!1.2$ Mpc.
The core of B1144+353 is a known variable radio source: its flux density at 1.4 GHz has increased continuously between 1974 and 1994. We have measured the flux density of the core in our WSRT observations (epoch 1997.7) and find a value of $541\pm10$ mJy. This implies that its flux density has decreased by $\sim\!70$ mJy between 1994 and 1997.
Further, we have retrieved unpublished archival ROSAT HRI data of B1144+352. The source has been detected and appears to be slightly extended in X-rays. We find a luminosity of $(1.26 \pm 0.15)\times10^{43}$ erg\,s$^{-1}$ between 0.1 and 2.4 keV, assuming that the X-ray emission is due to an AGN with a powerlaw spectrum with photon index 1.8, or $(0.95 \pm 0.11) \times10^{43}$ erg\,s$^{-1}$ if it is due to thermal bremsstrahlung at $T=10^7$~K. The detection of the X-ray source suggests that the intrinsic H{\sc i} column density cannot be much larger than a few times $10^{21}$~cm$^{-2}$. 
The non-detection of an extended X-ray halo in a radius of 250 kpc around the host galaxy limits the X-ray luminosity of an intra-cluster gas component within this radius to $\la2.3 \times 10^{42}$ erg\,s$^{-1}$ (1$\sigma$ upper limit). This is below the luminosity of an X-ray luminous cluster and is more comparable to that of poor groups of galaxies. Also the optical data show no evidence for a rich cluster around the host galaxy.
B1144+352 is the second GPS galaxy known to be associated with a Mpc-sized radio source, the other being B1245+676. We argue that the observed structure in both these GPS radio sources must be the result of an interrupted central jet-activity, and that as such they may well be the progenitors of sources belonging to the class of double-double radio galaxy.
\end{abstract}


\section{Introduction}

Gigahertz Peaked Spectrum (GPS, e.g. Spoelstra et al. 1985) radio sources are characterized by their convex radio spectra peaking at frequencies near 1 GHz. 
At frequencies above the peak, the radio spectrum has a power-law shape which is typical of an optically thin synchrotron emitting radio source (e.g. Pacholczyk 1970).
The turn-over of the radio spectrum at low frequencies can be due to Synchrotron Self Absorption (SSA) in high density parts of a compact radio source (e.g. Scott \& Readhead 1977), free-free absorption in a sheet of clouds surrounding the radio source (Bicknell, Dopita \& O'Dea 1997) or ionized clouds within it (Begelman 1998).\\

VLBI observations (e.g. Dallacasa et al. 1995, Stanghellini et al. 1997, Snellen 1997) show that GPS sources are compact radio sources with linear sizes up to a few hundred pc and often with a symmetrical double radio structure (Compact Symmetric Objects, CSO, e.g. Wilkinson et al. 1994, Readhead et al. 1996).
There are two popular scenarios to explain the origin of GPS sources. The first is that GPS sources are very young radio sources, that have just started to develop their radio lobes (Phillips \& Mutel 1982). They are expected eventually to grow into large double-lobed radio sources, perhaps even into the Mpc-sized Giant Radio Galaxies (GRGs). Because small radio sources are much more numerous than large radio sources, strong negative luminosity evolution as they increase in size is required (e.g. Fanti et al. 1995, Readhead et al. 1996, O'Dea \& Baum 1997). Reynolds \& Begelman (1997) propose that this can be achieved by intermittency of the jet-production in the AGN on timescales of $10^4 - 10^5$ yr.\\

The second scenario explains GPS sources as radio sources which are confined by a high density ISM (`frustrated' radio sources, e.g. van Breugel et al. 1984, O'Dea et al. 1991). In this scenario, the jet is not able to escape the dense surrounding material within its lifetime, either because it is repeatedly reflected off dense clouds or because the ram-pressure of the surrounding medium is high enough to practically stall the progress of the head of the jet. The radio source is therefore confined to a small central region.\\
   
Roughly 10\% of known GPS sources show extended (kpc-scale) radio emission (Baum et al. 1990, Stanghellini et al. 1990, O'Dea 1998). Baum et al. (1990) propose that the GPS source in these sources may either result from a disruption of an ongoing jet flow (`smothered' jets), or from recurrent nuclear radio activity. In both these models, the central (GPS) radio sources are young and the extended radio emission is older and fading since it is disconnected from the jet flow.
A convincing example of such a source is B0108+388, where motions of the radio components, observed with multi-epoch VLBI observations, imply that the pc-scale radio structure cannot be much older than $\sim\!350$ yr (Owsianik et al. 1998). The observed velocity of these components ($0.394 \pm 0.052c\,h_{50}^{-1}$) strongly suggests that their outflow is not hampered by a high density medium. However, Carilli et al. (1998) observe strong redshifted H{\sc i} absorption towards B0108+388, which, contrary to the expectation, suggests that the source actually is surrounded by a dense environment. B0108+388 also exhibits a faint kpc-scale radio structure (Baum et al. 1990, Carilli et al. 1998). This is considered as a strong indication that the nucleus is recurrently radio active.\\

Sources such as B0108+388 strongly suggest that some radio sources have recurrent phases of radio activity. Outstanding questions are how common recurrence is, and on what typical timescales it takes place? To answer these questions, it is important to find more of these apparently rejuvenated sources. A low-frequency search for extended radio emission on kpc-scales or larger around GPS/CSS sources should be able to accomplish this goal.\\

In this paper, we present new radio data of the region centered on the GPS source B1144+352, obtained with the Westerbork Synthesis Radio Telescope (WSRT) at 1.4~GHz, in addition to available radio data from the WENSS, NVSS and FIRST surveys. The WSRT radio map shows that B1144+352, apart from being a GPS source, also appears to be the host galaxy of a 1.2 Mpc large radio source. We further present unpublished ROSAT HRI data of the host galaxy, which shows it to be a luminous X-ray source.
To calculate physical properties we adopt values of $H_0 = 50$\,km\,s$^{-1}$\,Mpc$^{-1}$ and $q_0 = 0.5$\, throughout this paper. We define a radio spectral index $\alpha$ as $S_{\nu}\!\propto\!\nu^{\alpha}$.

\section{The GPS source B1144+352}
\label{sec:1144.intro}

The GPS nature of the radio source B1144+352 was first recognized by Snellen et al. (1995, hereafter S95), who find that the radio spectrum peaks at 2.4 GHz, with a spectral index of $+0.35$ in the optical thick (low frequency) part of the radio spectrum, and a spectral index of $-0.56$ in the optical thin part. These values should be treated with care, however, because of the variability of the central source (see also Sec. \ref{sec:1144_variability}) and the large time spanned by the observations used in obtaining them.\\
 
\begin{figure}
\resizebox{\hsize}{!}{\epsfig{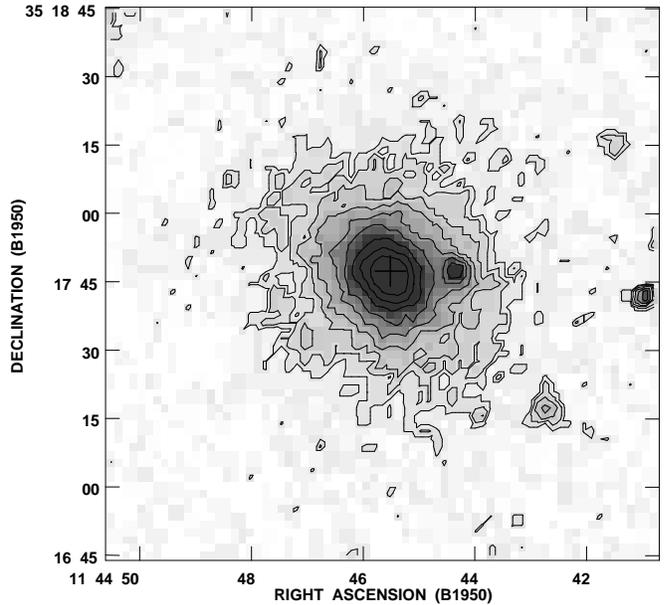}}
\caption{\label{fig:1144_dss} Contourplot of the host galaxy of the GPS source B1144+352 from the DSS. The contours are drawn at logarithmic intervals of $\sqrt2$, starting at three times the background noise level. The cross at the center of the host galaxy gives the position of the GPS source. Its size is not related to the positional uncertainty which is only $\sim\!0.5\arcsec$.}
\end{figure}

The radio source has been identified with the galaxy CGCG\,186-048 (Zwicky et al. 1960-1968). The Gunn $r$-band magnitude of the host galaxy is $14.3 \pm 0.1$ (Snellen et al. 1996) and it has a redshift of $0.0630$ (Colla et al. 1975a; Hewitt \& Burbidge 1991). This makes it one of the nearest GPS galaxies known. The absolute magnitude in the Gunn r-band is $-23.8 \pm 0.1$, using the K-correction as given in Snellen et al. (1996). The optical spectrum shows bright narrow emission lines, typical for powerful radio sources, on top of a starlight dominated continuum (Marcha et al. 1996). There is no hint of a non-thermal continuum or of broad Balmer emission lines, hence the classification as a GPS galaxy. An optical R-band CCD image can be found in Snellen et al. (1996). Fig. \ref{fig:1144_dss} presents a contour plot of the host galaxy from the Digitized Sky Survey (DSS). The host galaxy appears to have a small companion galaxy at a projected distance of $\sim\!25$ kpc (14.8\arcsec\,) to the west. This is close to the median distance at which apparent companions were found in a sample of GPS galaxies by O'Dea et al. (1996a). From the DSS it is apparent that CGCG\,186-048 is the brightest galaxy within a radius of at least 1~Mpc ($\sim\!20\arcmin$\,). There is no indication that it is a member of a rich cluster, although there are several fainter (R-band magn. $\ga15$) galaxies within a radius of a few hundred kpc. More likely, it is situated in a poor group of galaxies.\\ 

VLA maps of the GPS source at 1.4 GHz are presented by Parma et al. (1986) and S95. Both were obtained using the same configuration and have a resolution of $\sim\!4$ arcsec. The GPS source appears as a strong unresolved point-source. However, the maps also show a $\sim\!20\arcsec$ long weak jet-like feature emanating from the radio core, and a shorter ($\sim\!10\arcsec$) and somewhat weaker counter-jet. The total linear size of this extended system is $\sim\!65$ kpc. S95 note that the flux density from the 20-cm Green Bank survey (White \& Becker 1992) is higher (by $\sim\!80$ mJy) than that from their VLA observations. Since the Green Bank survey has a beamsize of $\sim\!10\arcmin$ (FWHM), they attribute this difference to a radio source component extending even beyond the observed jet-like feature. The fractional polarization of the core, obtained from high resolution 8.4-GHz VLA observations, is 3\% at a $2\sigma$-level (Marcha et al. 1996).\\

A VLBI map of the central source is presented by Henstock et al. (1995). It shows a $\sim\!40$ pc large double radio structure, with one well resolved bright radio lobe-like component and an unresolved second component. The position angle of the radio axis is $+121^{\circ}$, counted counter-clockwise (CCW) from the North. Giovannini et al. (1995) find that the two components are separating from each other superluminally with an apparent velocity of $2.4c\,h_{50}^{-1}$. This is surprising since GPS galaxies in general do not show superluminal motion (Stanghellini et al. 1997).\\

Giovannini et al. (1990) report that the flux density of the peak in the radio spectrum (the peak flux density) has continuously increased from $\sim\!300$~mJy in 1974 to $\sim\!600$~mJy in 1990. S95 and Snellen et al. (1998) claim that it has decreased again somewhat thereafter. Snellen et al. (1998) further state that the brightest (eastern) VLBI component, which has a radio lobe-like morphology, must be largely responsible for the observed change in the peak flux density. Variability in the flux density has been found in other GPS sources, but mostly in GPS quasars (e.g. Stanghellini et al. 1998).\\

\section{New observations}

\begin{figure}
\resizebox{\hsize}{!}{\epsfig{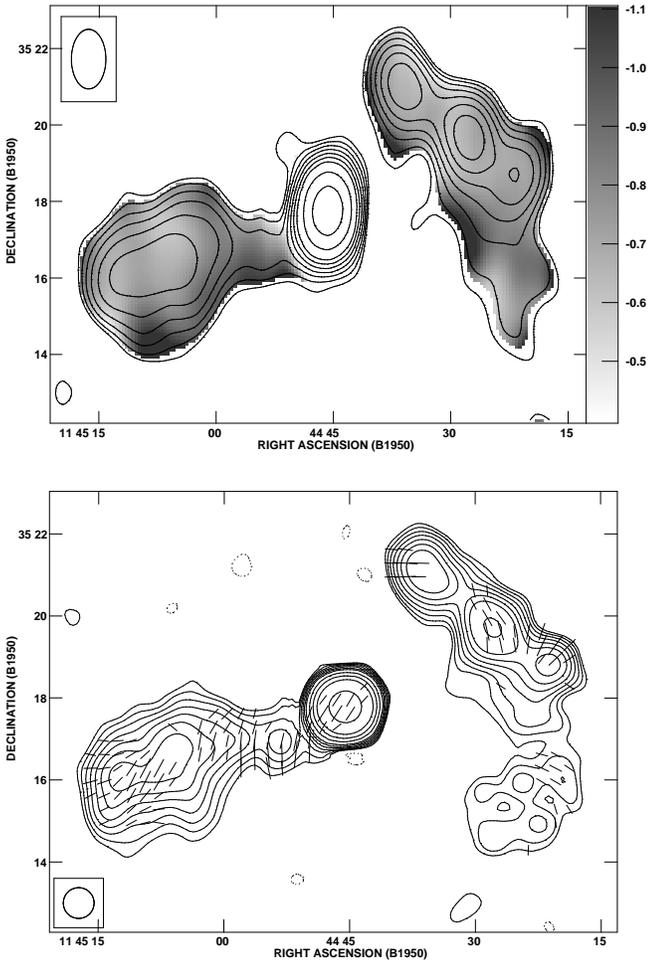}}
\caption{\label{fig:1144.wenss+nvss}Radio maps of the source B1144+352 from the WENSS and the NVSS surveys. {\bf a} Contour plot from the 325-MHz WENSS survey. Contourlevels are at ($-8$,8,11.3,16,22.6,32,45.3,64,128,256,512) mJy beam$^{-1}$. The grey scale represents the spectral index $\alpha$ between 1400 and 325~MHz, calculated using only pixels above a $4\sigma$-level in the WENSS and convolved NVSS maps. It ranges from $-0.4$ (white) to $-1.1$ (black). {\bf b} Contour plot from the 1.4-GHz NVSS survey. Contours are at ($-1.2$,1.2,1.7,2.4,3.4,4.8,6.8,9.6,19.2,38.4,76.8,307.2) mJy beam$^{-1}$. Also plotted are the observed $E$-field polarization vectors, with a length proportional to the polarized intensity (1\arcsec $=0.05$ mJy beam$^{-1}$).}
\end{figure}

\subsection{WENSS, NVSS and FIRST radio survey data}

B1144+352 has been observed in three recent major sky surveys: WENSS (WSRT 325 MHz, Rengelink et al. 1997), NVSS (VLA D-array 1.4 GHz, Condon et al. 1998) and FIRST (VLA B-array 1.4 GHz, Becker et al. 1995). The WENSS radio map, with a (FWHM) beamsize of $54\arcsec ({\rm R.A.}) \times 94\arcsec ({\rm Dec.})$ at the declination of B1144+352, shows the bright unresolved GPS radio core and two diffuse extended radio structures on either side of it (see Fig. \ref{fig:1144.wenss+nvss}a). The eastern structure is connected to the GPS source by a faint `bridge' and has a morphology which resembles the lobes of edge-brightened (or FRII-type, Fanaroff \& Riley 1974) double-lobed radio sources. The western structure is elongated in a direction perpendicular to the radio axis as defined by the core and the eastern lobe. It consists of three bright regions and an extended, diffuse southern tail.\\
 
The 1.4-GHz NVSS survey, which uses a beamsize of 45\arcsec~(FWHM), has also detected the diffuse extended structures and the southern tail (see Fig. \ref{fig:1144.wenss+nvss}b). The total flux density of the extended radio structure in the NVSS maps is $247 \pm 8$~mJy, which is much more than the 80~mJy discrepancy that S95 found between their core flux density and the flux density found in the 20-cm Green Bank survey. We believe that there are two reasons for this difference. 
First, the Green Bank survey was conducted in April 1983, when the flux density of the core was much lower than during the observation of S95 (see Sect. \ref{sec:1144_variability} and Tab. \ref{tab:1144_coreflux}). Second, the flux density given in White \& Becker (1992) is the peak flux density, which, for an extended source like B1144+352, is an underestimate of the total flux density.\\

We have used the NVSS and the WENSS maps to produce a spectral index map, which is shown in grey scale in Fig. \ref{fig:1144.wenss+nvss}a. To avoid artefacts resulting from combining data with different beamsizes, we have first convolved the NVSS map to the resolution of WENSS. The `bridge' that connects the central source to the eastern structure has a somewhat steeper radio spectrum than the rest of that structure. The surface brightness of the tail of the wetsern structure is rather low, so that the spectral index is poorly determined. We measure a value of $\sim\!-0.85$, which is not much different from the other extended source components.\\

The NVSS has further detected linear polarization in both extended radio structures. Halfway through the eastern lobe, the position angles of the vectors of the $E$-field of the linear polarized emission change from perpendicular to more parallel with respect to the radio axis. Polarization has also been detected towards the core. The fractional polarization is $0.4\pm0.1$\%. It will be shown in Sect. \ref{sec:1144_wsrt} that this polarized emission mainly originates in the eastern jet and not in the GPS radio core.\\

\begin{figure*}
\resizebox{\hsize}{!}{\epsfig{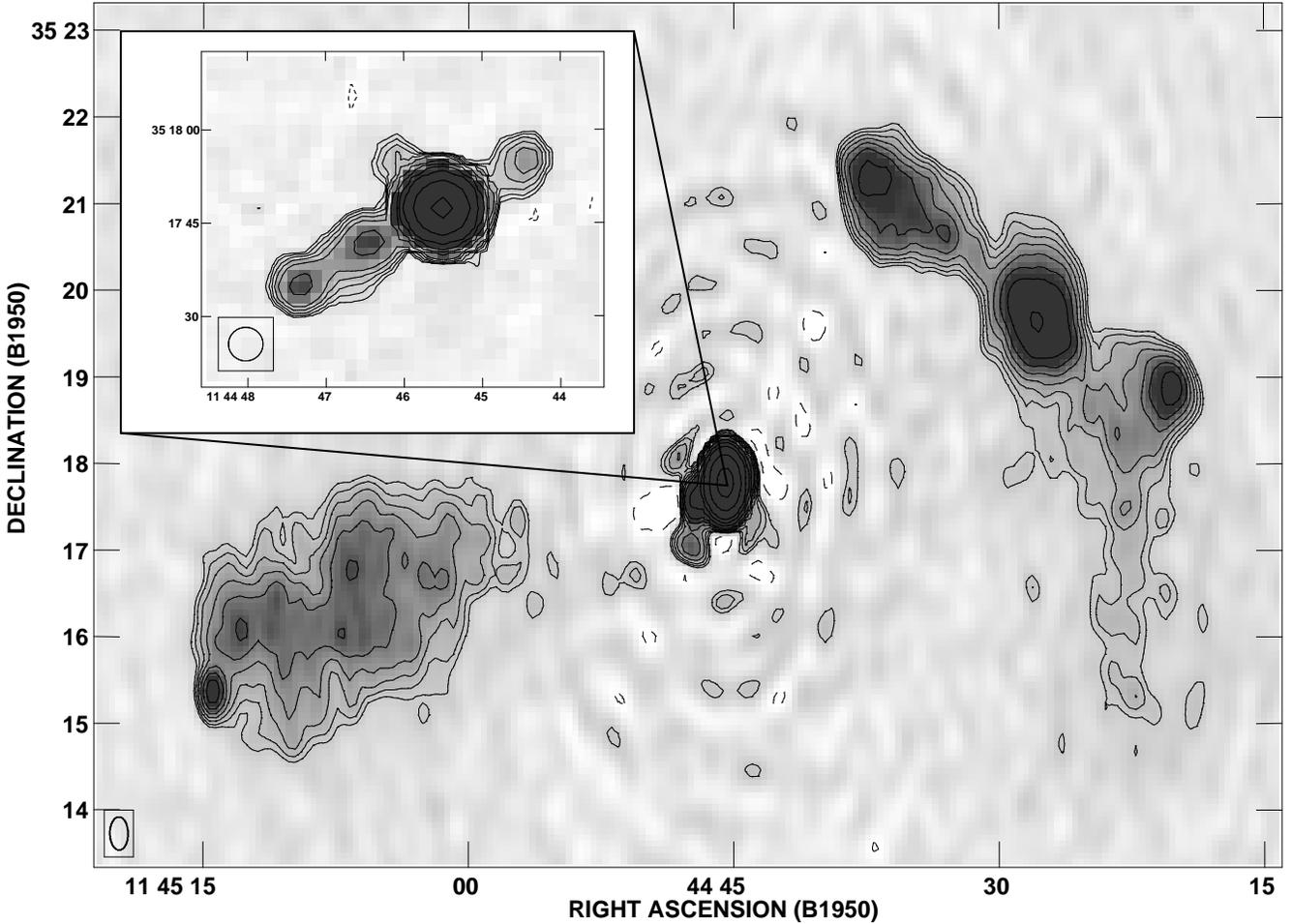}}
\caption{\label{fig:1144.wsrt+first}Radio map of the source B1144+352 from our 1.4-GHz WSRT observations. The contours are at ($-0.3$,0.3,0.42,0.6,0.85,1.2,1.7,2.4,4.8,9.6,19.2,76.8,307.2) mJy beam$^{-1}$. The greyscale ranges from $-0.3$ to 2.0 mJy beam$^{-1}$ and clearly shows the ringlike artefacts in the radio map. The inset is a radio map of the central source from the FIRST survey. Contours are at ($-0.45$,0.45,0.64,0.90,1.27,1.8,3.6,7.2,14.4,28.8,115.2,460.8) mJy beam$^{-1}$.}
\end{figure*}

The radio source has also been mapped in the FIRST survey, which uses a beamsize of 5\farcs4 (FWHM). Therefore it has a slighter lower resolution than the 1.4-GHz VLA B-array observations of Parma et al. (1986), but it has a comparable sensitivity. The FIRST radio map (see the inset in Fig. \ref{fig:1144.wsrt+first}) confirms the presence of the jet and counter-jet like features. The position of the core, determined by fitting a Gaussian to the central component, is $11^h\,44^m\,45\fs52$ in Right Ascension and $+35\degr17\arcmin47\farcs4$ in Declination (B1950.0), with a positional uncertainty of $\sim\!0.5\arcsec$. The radio axis of the structure visible in FIRST and the map of Parma et al. (1986) has the same position angle ($+120\degr$, measured CCW from the North) as that of the pc-scale structure mapped by Henstock et al. (1995). There is therefore no indication of a change in the jet axis between pc and kpc-scales. A summary of the properties of the central component, as measured in the FIRST radio map, is presented in Tab. \ref{tab:1144_first}.

\begin{figure*}
\resizebox{12cm}{!}{\epsfig{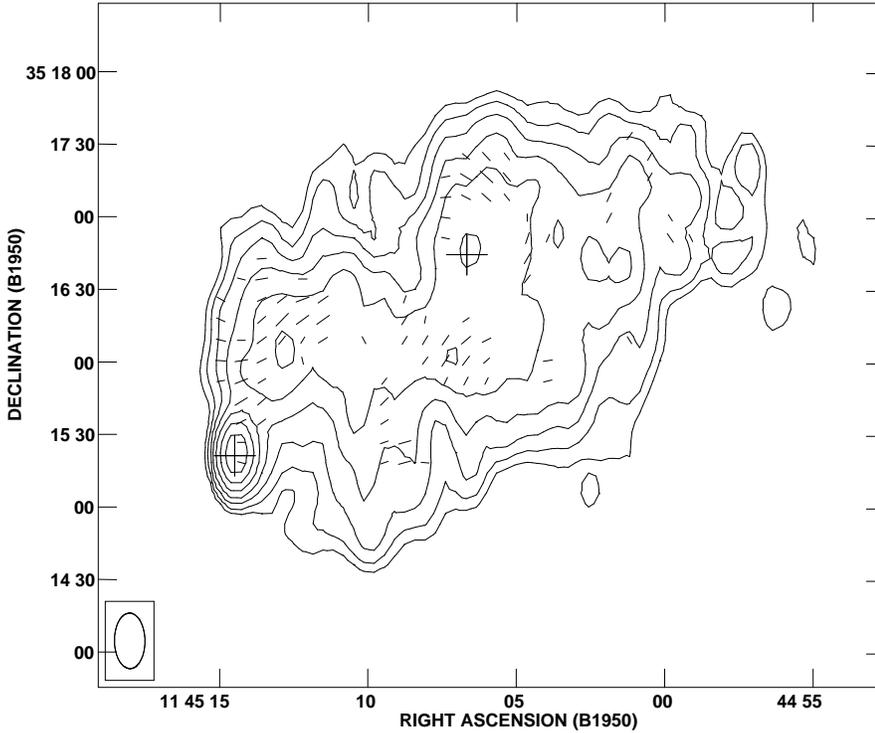}}
\hfill
\parbox[b]{60mm}{
\caption{\label{fig:1144.east} Contourplot of the eastern radio structure from our 1.4-GHz WSRT observations. Contours are at ($-0.3$,0.3,0.42,0.6,0.85,1.2,1.7,2.4) mJy beam$^{-1}$. The vectors are the observed $E$-field, and their length corresponds to the polarized intensity ($1\arcsec = 0.05$ mJy beam$^{-1}$). The two crosses give the position of sources detected in the FIRST survey; their size is arbitrary and not related to any physical quantity or the positional accuracy of these sources.}}
\end{figure*}

\begin{table}
\caption{\label{tab:1144_first}Properties of the central component of B1144+352 as measured in the FIRST survey map.}
\begin{tabular}{l r@{$\pm$}l l l}
\hline
Comp. & \multicolumn{2}{c}{$S_{1435}$} & \multicolumn{1}{c}{$D_{\rm max}$} & \multicolumn{1}{c}{$PA$} \\
& \multicolumn{2}{c}{$[$mJy$]$} & \multicolumn{1}{c}{$[$kpc$]$} & \multicolumn{1}{c}{$[\,\degr\,]$} \\
\hline\\
Core & 609.3 & 0.3 & & \\
`Jet' & 14.5 & 0.5 & 41.2 & \php119.9 \\
`Counter-jet' & 3.2 & 0.4 & 24.5 & $-59.5$ \\
\hline
\end{tabular}
\vskip 1mm
Notes: The jet is the eastern component, the counter-jet the western. The flux density at 1435~Mhz, $S_{1435}$, of the core is the peak flux determined using a Gaussian fit to this component on the map. The date of the FIRST observations is 3 July 1994. The flux densities of the jet and the counter-jet have been measured on the map after subtraction of the core. $D_{\rm max}$ is the projected distance between the most outer peak in flux density of the jet/counter-jet and the core. The position angles, $PA$, are measured counter-clockwise from the North. 
\end{table}

\subsection{1.4-GHz WSRT observations}
\label{sec:1144_wsrt}

To better understand the nature of the extended structures around the GPS source, we observed B1144+352 for 12 hr with the WSRT at 1.4 GHz on August 31, 1997. We used a total bandwidth of 50~MHz, divided into five channels of 10~MHz each and centered on 1395~MHz. 11 out of 14 antennas were operational during these observations. The sources 3C\,286 and 3C\,147 were used as primary calibrators. 
The data have been reduced and mapped using the NFRA data reduction package {\sc newstar}. We used the flux density scale of Baars et al. (1977) for absolute gain calibration. During the reduction process, we encountered a problem in the data, manifested as a system of ringlike structures around the bright central source (see Fig. \ref{fig:1144.wsrt+first}). Our best guess is that it is due to some kind of time-variable baseline-based errors.  
The radial decrease of this effect ensures that the observed structure of the diffuse lobes is not influenced much by it.\\

\begin{figure}
\resizebox{\hsize}{!}{\epsfig{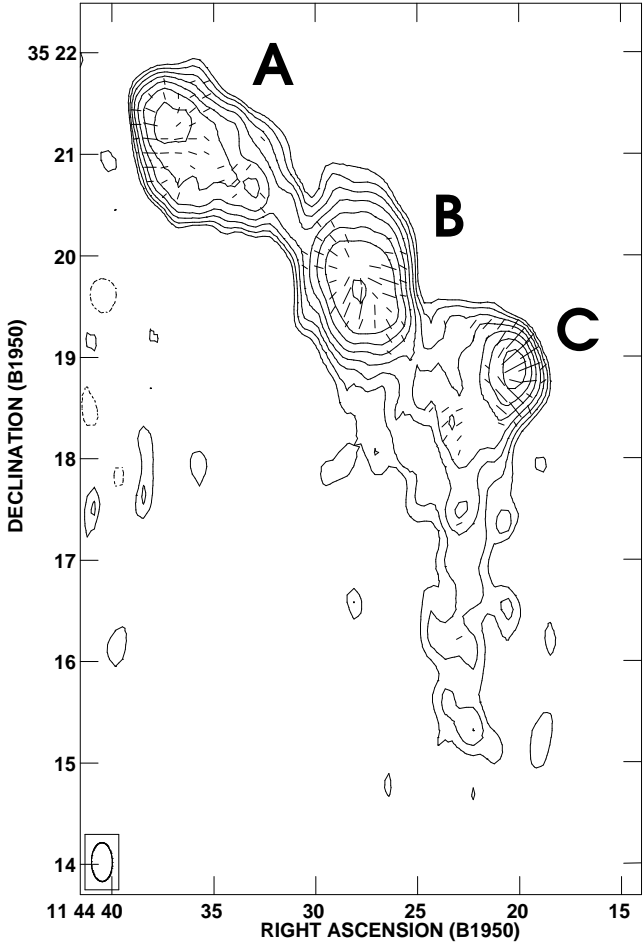}}
\caption{\label{fig:1144.west} Contourplot of the western radio structure from our 1.4-GHz WSRT observations. Contours are at ($-0.3$,0.3,0.42,0.6,0.85,1.2,1.7,2.4,4.8) mJy beam$^{-1}$. The vectors represent the observed $E$-field of the linear polarization; their length corresponds to the polarized intensity ($1\arcsec = 0.05$ mJy beam$^{-1}$).}
\end{figure}

The final total power map has an {\sc rms} noise of 0.053~mJy, although this value increases substantially towards the center of the map. The FWHM beamsize of the restoring beam is $12\farcs58\,({\rm R.A.})\!\times\!22\farcs97\,({\rm Dec.})$. The Stokes' $Q$ and $U$ parameter maps have the same {\sc rms} noise as the total power map. Only the $Q$-map has similar ringlike structures to those seen in the total power map.
Both diffuse structures have been detected and mapped. The FRII-type radio lobe morphology of the eastern structure (see also Fig. \ref{fig:1144.east}) is obvious, and we also detect a leading hotspot. 
Surprisingly, the western structure (see also Fig. \ref{fig:1144.west}) appears to be a superposition of a separate FRII-type radio source and a radio lobe with a tail towards the south. The position angles of the $E$-field vectors of the linear polarized emission in the two extended structures (Fig. \ref{fig:1144.east} and Fig. \ref{fig:1144.west}) are consistent with those measured in the NVSS (Fig. \ref{fig:1144.wenss+nvss}b). Near the core, we only measure significant polarization at the position of the most eastern component of the FIRST radio map. The polarized intensity is $1.7 \pm 0.4$ mJy, and the fractional polarization of this component is therefore $35 \pm 9$\%.  Integrated over the whole central component, the fractional polarization is $0.3\pm0.1$\%, which is equal to what we found in the NVSS data. Towards the radio core no significant polarized emission is detected. The position angle of the $E$-field vector of the linear polarized emission is $+123\degr$, and so the observed projected $E$-field is parallel with the radio axis. 

\begin{figure}
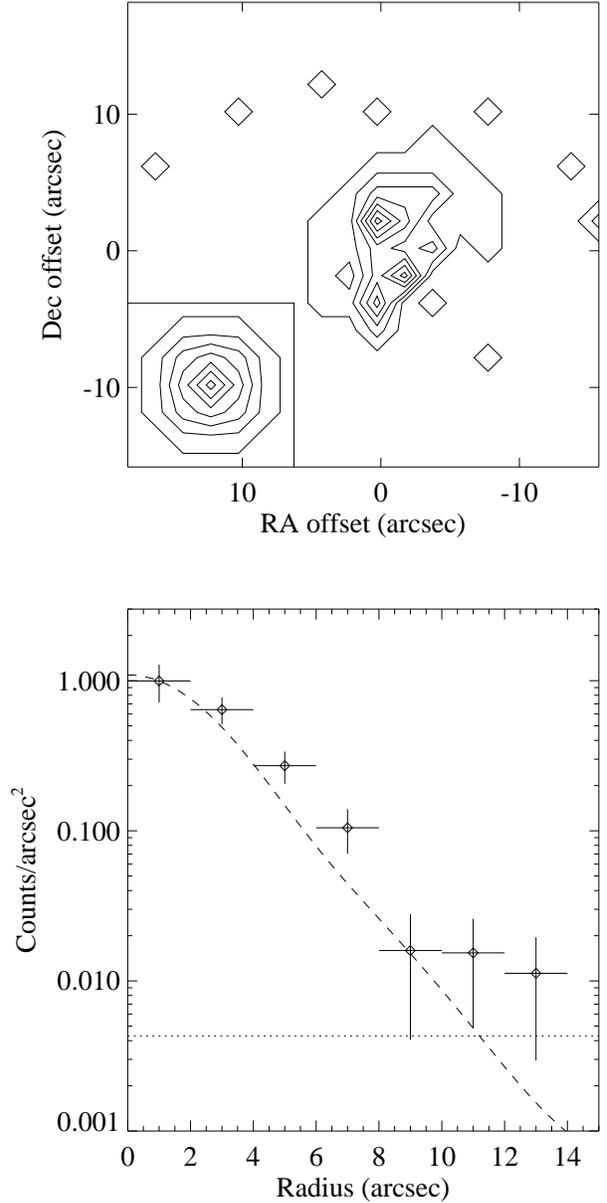

\resizebox{\hsize}{!}{\epsfig{file=MS8045.f6a}} \\
\resizebox{\hsize}{!}{\epsfig{file=MS8045.f6b}}
\caption{\label{fig:1144_rosat}{\bf a} Contourplot representation of the raw counts in the ROSAT HRI data in the area surrounding the X-ray source associated with B1144+352. The counts have been binned on a grid with 2 arcsec large cells. Contourlevels are at $0.25 \times $(1,2,3,4,6,7) counts arcsec$^{-2}$. The coordinates are relative to the peak of the central radio component visible in the FIRST survey. The inset in the lower left corner shows the HRI point response function, scaled to the peak flux of the image and using the same contourlevels. {\bf b} Plot of the radial profile of the raw ROSAT counts. The dashed line is the HRI point response function, normalized to the counts in the innermost bin. The dotted line is the background level, measured in an annulus from 200 to 800 arcsec from the central source.}  
\end{figure}

\subsection{ROSAT observations}

B1144+352 has also been observed in X-rays by ROSAT on 10 June 1993. We retrieved the X-ray data, obtained with the High Resolution Imager (HRI, David et al. 1993), from the ROSAT archive. The X-ray image available from the archive, which has been binned with 8 arcsec pixels (the FWHM of the HRI is $\sim\!5\arcsec$), shows a rather strong point-like source, whose position, within the ROSAT pointing errors, agrees well with that of the central radio source in the FIRST radio map ($\Delta{\rm RA} =0\farcs8$, $\Delta{\rm Dec}=0\farcs5$). In this 3000\,s exposure, $70\pm8$ net counts have been detected from this source. Two other sources are visible near the edge of the field, but there are no other significant detections within the extent of the radio source.\\

To investigate the structure of the X-ray source in more detail we rebinned the original data on a grid with 2 arcsec large cells. The result is shown in Fig. \ref{fig:1144_rosat}a, which is a contourplot of the raw datacounts. We have set the coordinate frame such that after convolution with an 8\arcsec (FWHM) Gaussian the position of the peak corresponds to the peak of the radio emission from the FIRST survey. For comparison, we have also plotted the ROSAT HRI point response function (PRF\footnote{As function of radius $R$ given by $PRF(R) = A_1 \exp(-0.5 (R/S_1)^2) + A_2\exp(-0.5 (R/S_2)^2) + A_3\exp(-R/S_3)$, with $R$ in arcsec, $A_1 = 0.9638, A_2 = 0.1798, A_3 = 0.0009, S_1 = 2\farcs1858, S_2 = 4\farcs0419, S_3= 31\farcs69$ (David et al. 1993).}) after scaling it to match the peak intensity in the image.
The X-ray source appears to be extended in a direction close to the radio axis. This extension did not disappear if we shifted our binning grid by 1 arcsec in either direction. Further, the exposure was continuous, and therefore could not be hampered by pointing errors between different exposures. In Fig. \ref{fig:1144_rosat}b we have plotted the radial profile of the counts. The dashed line gives the PRF of the HRI. There is some evidence for extended structure in the inner 8 arcsec. However, a much longer exposure would be needed to firmly establish it.\\ 

To convert the X-ray counts to a received flux, we used the {\sc pimms} mission simulation program (version 2.4b; Mukai 1993). The X-ray flux can either be due to the AGN, in which case a powerlaw spectrum is expected, or to thermal bremsstrahlung in a shocked ISM. In case of B1144+352, the reality is probably a mixture of these. Therefore we calculate the X-ray luminosity using both a powerlaw and a thermal bremsstrahlung model and regard these as two limiting cases.  
For the powerlaw model we use a photon index of 1.8\,, which has been found to be the average for high-luminosity AGN (Williams et al. 1992). To correct for galactic extinction, a galactic atomic hydrogen column density of $1.84\times10^{20}$ cm$^{-2}$ towards B1144+352 has been used, as determined from the Leiden-Dwingeloo H{\sc i}-survey (Hartmann 1994). We have no information on the amount of extinction in or near the source, and so we neglect this contribution in our calculations.
We find a total X-ray flux density of $(8.1\pm1.0) \times 10^{-13}$ erg\,cm$^{-2}$\,s$^{-1}$ in the $(0.1-2.4)$ keV band. This translates to a total X-ray luminosity of $(1.26 \pm 0.15)\times10^{43}$ erg\,s$^{-1}$ between 0.1 and 2.4 keV. For the thermal bremsstrahlung model we use a gas temperature of $10^7$ K ($kT = 0.86$~keV) and a metal abundance of $0.25$ solar (model {\sc rsq70} in {\sc pimms}). Using the same galactic H{\sc i} column density as before, we find a total flux of $(6.1\pm0.7) \times 10^{-13}$ erg\,cm$^{-2}$\,s$^{-1}$ in the $(0.1-2.4)$ keV band. The total emitted power in this band then yields $(0.95 \pm 0.11) \times10^{43}$ erg\,s$^{-1}$.\\

Crawford \& Fabian (1995) find similar values for the four narrow-line radio galaxies at $z<0.1$ in their sample of powerful radio galaxies from the 3CR sample. Worrall \& Birkinshaw (1994) find an order of magnitude lower values for low-power radio galaxies. The X-ray luminosity of B1144+352 is therefore more comparable to that of powerful radio galaxies.
The detection of such a strong X-ray source in a GPS galaxy is surprising. O'Dea et al. (1996b) observed two GPS radio sources with the ROSAT PSPC and no X-ray emission was detected. Therefore the X-ray luminosity of these sources must be $\la 3 \times 10^{42}$ erg\,s$^{-1}$ in the $0.2-2$ keV band (3$\sigma$ upper limit), which is a factor of three below the luminosity of B1144+352. 
O'Dea et al. explain the non-detections by stating that these sources either must be intrinsically weak in X-rays and/or they must be highly obscured by cold gas. To obscure a source with an X-ray luminosity of $10^{43} - 10^{44}$ erg\,s$^{-1}$ in their observations, a column density of a few times $10^{22}$ cm$^{-2}$ is needed. The high X-ray luminosity of B1144+352 is therefore suggestive of a much lower column density towards this source: for the powerlaw model, a column density of a few $10^{22}$ cm$^{-2}$ would imply an intrinsic X-ray luminosity of $\sim\!10^{44}$ erg\,s$^{-1}$. 
However, a more modest column density of $5 \times 10^{21}$ cm$^{-2}$ yields an emitted power of $1.7\times10^{43}$ erg\,s$^{-1}$, which is only 35\% higher than the power that we find assuming no intrinsic absorption.\\
 
As will be shown in Sect. \ref{sec:radio_discussion}, we believe that the extended radio components are associated with the GPS source. This means that some medium must be present to confine the radio lobes. Such a medium may reveal itself through the emission of X-rays. However, the HRI data show no sign of a large extended X-ray emitting halo. To find an upper limit for the luminosity of such a halo we have summed all counts in a circular area of radius 250 kpc centered on the host galaxy of B1144+352. This is the typical core radius of low-luminosity clusters (Mulchaey et al. 1996; Jones et al. 1998) and therefore most of the X-ray halo emission is expected to come from this region. 
To estimate the background contribution we have summed all counts in a 600 arcsec wide annulus with an inner radius of 200 arcsec (330 kpc). In the inner circular area we measure 391 counts. The expected number of background counts in this region is $311 \pm 18$, and the inner 10\arcsec, or 15 kpc, contains $70 \pm 8$ counts. Therefore the number of counts from a possible X-ray halo in the HRI data, integrated from 15 kpc outwards to 250 kpc, is $10 \pm 26$ and thus not significant.  
To calculate an upper limit for the luminosity of the halo we have again used the {\sc pimms} mission simulation program (version 2.4b). We further assume that the density profile, $n(r)$, of the halo follows a modified King model, $n(r) = n_0 \times [1+(r/a)^2\,]^{-3\beta/2}$, with core radius $a=250$~kpc and $\beta = 0.75$ (e.g. Forman \& Jones 1991). The radial surface brightness profile then behaves as $I(r) = I_0 \times [1+(r/a)^2\,]^{0.5 - 3\beta}$, which implies that 10\% of the X-ray luminosity emitted innerhalf of 250 kpc originates from within a radius of 15 kpc. 
For the cluster gas, we have assumed a Raymond-Smith spectral model (model {\sc rsq70}) with a temperature of $10^7$~K (0.86 keV) and a metal abundance of 0.25 solar. These are reasonable values for the medium in groups of galaxies (Mulchaey et al. 1996). Using the 1$\sigma$ flux-level as an upper limit, we find that the total received flux in the $(0.1-2.4)$ keV band of the ROSAT HRI detector must be $\la\!2.25\times10^{-13}$ erg\,s$^{-1}$\,cm$^{-2}$. 
To compare this value with the literature we calculate the expected flux in the $(0.5-2$) keV band, which results in an upper limit of $1.9\times10^{-13}$ erg\,s$^{-1}$\,cm$^{-2}$. At the redshift of B1144+352, this translates into a 1$\sigma$ upper limit of the X-ray luminosity of $2.9 \times 10^{42}$ erg\,s$^{-1}$ in the inner 15 to 250 kpc. Corrected for the inner 15 kpc, this implies an upper limit of the X-ray halo luminosity within a radius of 250 kpc of $3.2 \times 10^{42}$ erg\,s$^{-1}$. Extrapolated to much larger radii, assuming the modified King profile, this would yield $4.3 \times 10^{42}$ erg\,s$^{-1}$.\\

The luminosity we find is much lower than those for luminous X-ray clusters ($10^{44} - 10^{45}$ erg\,s$^{-1}$) which is consistent with the lack of a rich optical cluster on the DSS.
Mulchaey et al. (1996) and Mulchaey \& Zabludoff (1998) have investigated the X-ray properties of a sample of nearby poor groups of galaxies. In the majority of cases, they find extended halo components with a typical luminosity range of $10^{42} - 10^{43}$ erg\,s$^{-1}$. They attribute this to a low-mass version of the intra-cluster medium found in X-ray luminous clusters. 
The upper limit we determined from ROSAT HRI data for B1144+352 therefore does not exclude a low-mass intra-cluster medium as found in poor groups of galaxies.  
   
\section{Results and Discussion}

\subsection{The radio light curve of the central source}
\label{sec:1144_variability}

\begin{figure}
\resizebox{\hsize}{!}{\epsfig{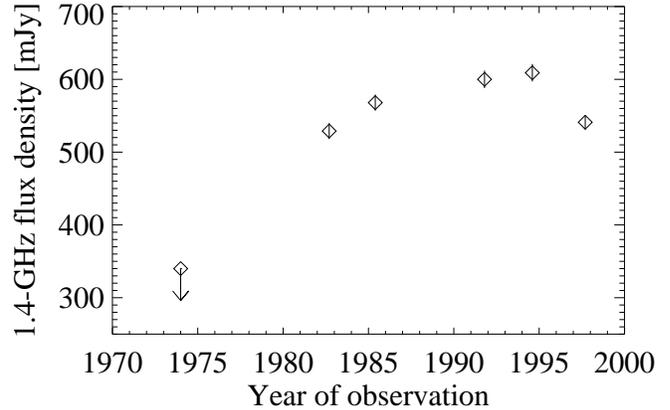}}
\caption{\label{fig:1144_coreflux} The 1.4-GHz radio light curve of the core of B1144+352 over the last 25 years. Values and references for the flux densities presented here can be found in Tab. \ref{tab:1144_coreflux}.} 
\end{figure}

We have measured the flux density of the central source at 1.4 GHz from the WSRT data by fitting a point-source model directly to the interferometer visibilities. This yields a flux density of $541 \pm 10$~mJy. We have tabulated and plotted several other 1.4-GHz flux density measurements from the literature in Tab. \ref{tab:1144_coreflux} and Fig. \ref{fig:1144_coreflux}.
The measurements from Parma et al. (1986), S95 and the FIRST survey have been made using the VLA in its B-configuration, and are therefore sensitive to the same source structures. Fanti et al. (1987) used the VLA in its A-configuration and therefore their observations have been done with a $\sim\!3$ times higher resolution. However, since the central source is unresolved by them, their flux density measurement should be comparable to the lower resolution ones. We have assumed a 2\% gain calibration error in all VLA observations.
The value from Colla et al. (1975a) is strictly an upper limit since this measurement has been done using the Nancay radio telescope with a beamsize of 4\arcmin~in R.A. and 24\arcmin~in Dec. (Colla et al. 1975b) and therefore also includes a fraction of the extended emission.\\

We find that, after a continuous rise between 1974 and 1994, the core flux density at 1.4 GHz has decreased by $\sim\!70$ mJy in the three years between the latest VLA observations (FIRST survey) and these observations.
Although several GPS sources have been found to be variable, they are usually quasars and not galaxies (e.g. Stanghellini et al. 1998). The (at least) 20 year long continuous rise and subsequent fall in the flux density of B1144+352 is therefore quite a remarkable behaviour.
Snellen et al. (1998) present a model for the flux density evolution of GPS sources, in which such a turn-over in flux density is expected. In their model, the radio components are ejected from the nucleus at relativistic velocities and at a large angle to the line of sight. As a result of relativistic beaming, the intensity of the radiation emitted by such a component in the direction of the observer will be very low. 
Only after the radio component has decelerated sufficiently, it will start to contribute significantly to the observed flux density. While the component is decelerating, it will also expand adiabatically, so that the total emitted flux will drop. The exact behaviour of the observed flux density depends on the relative importance of these two effects during the evolution of the radio source.
Snellen et al. (1998) show that during the initial phases of the deceleration the flux density will increase, due to the declining effect of relativistic beaming, while after some time, as a result of adiabatic expansion of the component, the flux density will slowly decrease. According to their model, the eastern lobe visible in the VLBI map of Henstock et al. (1995), which is largely responsible for the observed change in flux density, must be in the expansion phase. This can be tested by monitoring the peak frequency, which should now also be decreasing in time.    

\begin{table}
\caption{\label{tab:1144_coreflux}1.4-GHz flux densities of the core of B1144+353.}
\begin{tabular}{l r@{$\pm$}l l}
\hline
Year & \multicolumn{2}{c}{Flux density} & Ref. \\
 & \multicolumn{2}{c}{$[$mJy$]$} & \\
\hline
1974 & 340 & 34 & C75 \\
1982.7 & 529 & 11 & P86 \\
1985.4 & 568 & 11 & F86 \\
1991.8 & 600 & 12 & S95 \\
1994.6 & 609 & 12 & F \\
1997.7 & 541 & 10 & W \\
\hline \\
\end{tabular} 
\vskip 1mm
Notes: Column 1 refers to the epoch of observation. Column 2 gives the flux density at 1.4 GHz. Column 3 gives the reference for the flux density measurement (C75: Colla et al. 1975a; P86: Parma et al. 1986; F87: Fanti et al. 1987; S95: Snellen et al. 1995; F: FIRST survey; W: This paper).
\end{table}

\subsection{The large scale radio structure}
\label{sec:radio_discussion}
 
Although we do not detect radio jets to physically connect the central GPS source with the diffuse outer structures, there are several indications that they are truly associated, and that the GPS host galaxy is therefore also the host galaxy of a co-aligned Mpc-sized radio source. In this section, we will look at the evidence for each radio structure separately.
 
\subsubsection{The eastern radio structure} 

The eastern structure (Fig. \ref{fig:1144.east}) resembles an extended FRII-type radio lobe with a leading hotspot and with an overall morphology that points back to the GPS source. The hotspot is also detected in the FIRST survey, with a flux density of $3.03 \pm 0.15$ mJy. There is also a second weak ($1.36\pm0.15$~mJy) radio source in the FIRST survey at R.A. 11$^h$\,45$^m$\,06\fs68, Dec. $+35\degr$\,16\arcmin\,44\farcs8, which coincides with a local maximum in the radio emission in the WSRT map. The positions of these two sources have been marked with a cross in Fig. \ref{fig:1144.east}.\\

The detection of these sources provides two alternative explanations for the origin of this radio structure. First, it could be an unrelated head-tail radio source with the hotspot actually being the radio core of such a structure.
If this were the case then we would expect to find an optical galaxy coinciding with the FIRST radio source. However, there is no object brighter than an R-band magnitude of 20 at this position in the Digitized Sky Survey (DSS). If the host galaxy of this radio source would be in the same group as the GPS host galaxy, its non-detection implies that its absolute magnitude is at least 6 mag. weaker in R-band than that of the GPS host (i.e. $M_R \ga -17.8$). It is unlikely that such a weak galaxy would harbor an AGN capable of producing the observed radio structure (see, e.g., Owen 1993).  
The other possibility is that the host galaxy is at a much higher redshift and that the observed position on the sky near the GPS source is just a coincidence. This would imply that the head-tail source is extremely large and powerful. 
Also, if this radio structure were a head-tail source, one would expect a significant steepening of the radio spectrum towards the end of the tail. We do not observe such a radio spectral behaviour (see Fig. \ref{fig:1144.wenss+nvss}a). Therefore, we regard it unlikely that the eastern radio structure is a separate head-tail radio galaxy.\\

Second, the eastern radio structure could be a somewhat amorphous radio source with, possibly, the second weak radio source detected by FIRST in the middle of this structure as the radio core.   
Again, the lack of an optical counterpart in the DSS, situated within the boundaries of the eastern radio structure, strongly argues against this scenario.\\ 
 
We can therefore rule out the possibility that the eastern radio structure is a separate and unrelated radio source, leaving only the conclusion that it must be associated with the GPS source. Both the NVSS and the WENSS radio maps show a faint `bridge' connecting the eastern radio structure with the GPS source. The WSRT map does not show this bridge, but that is most likely related to the problem in the data discussed in Sect. \ref{sec:1144_wsrt}. The detection of this bridge is further evidence that the eastern radio structure is a radio lobe originating from within the GPS host galaxy. 

\begin{figure}
\resizebox{\hsize}{!}{\epsfig{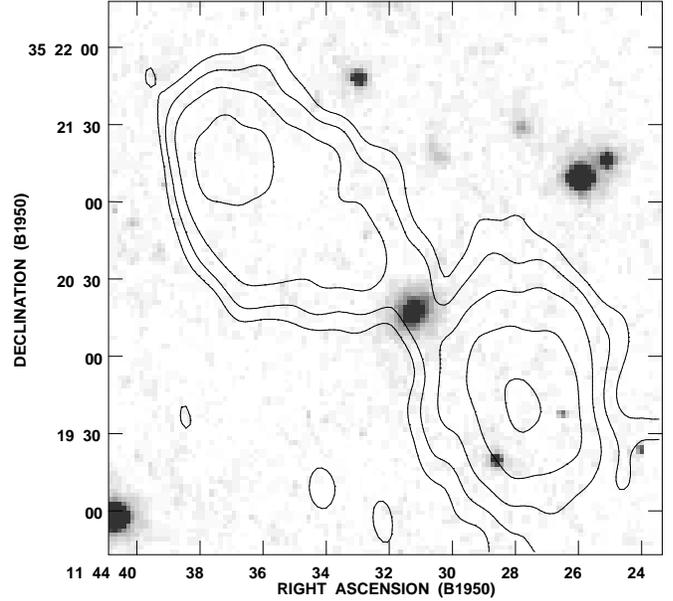}}
\caption{\label{fig:B1144+353.dss} Contour plot of our 1.4-GHz WSRT observations of radio components A and B, overlaid on a greyscale plot of the optical field from the DSS. Note the presence of a bright galaxy between the two radio components. Contourlevels are at (0.3,0.6,1.2,2.4,4.8,9.6,19.2) mJy beam$^{-1}$.}
\end{figure}

\begin{figure}
\resizebox{\hsize}{!}{\epsfig{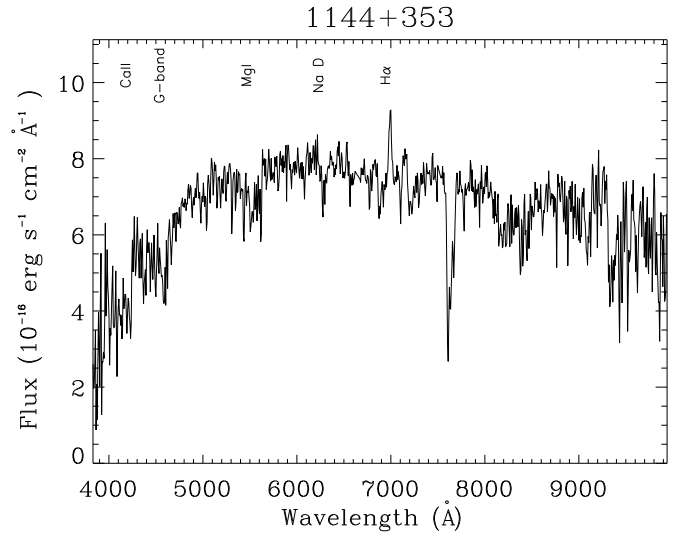}}
\caption{\label{fig:B1144+353.spectrum} Optical spectrum of the bright galaxy situated between components A and B (see Fig. \ref{fig:B1144+353.dss}). The features we have used to determine the redshift are indicated. There is weak H$\alpha$+$[$N{\sc ii}$]$ emission ($\rm{EW}=10\pm1$\AA).~The strong absorption feature near 7600\AA~is atmospheric.}
\end{figure}

\begin{table}
\caption{\label{tab:B1144+353}Properties of the radio source B1144+353.}
\begin{tabular}{l l}
\hline \\
Host galaxy R.A. & 11$^h$44$^m$31\fs25\\
Host Galaxy Dec. & $+35$\degr20\arcmin17\farcs5\\
$R$-band magn. & $15.3 \pm 0.5$\\
Redshift & $0.065 \pm 0.001$ \\
Linear size & 350 kpc \\
Flux density & $70 \pm 3$ mJy \\
Radio power & $1.3 \times 10^{24}$ W\,Hz$^{-1}$ \\
$B_{eq}$ & 1.2 $\mu$G \\
$u_{eq}$ & $1.4 \times 10^{-13}$ erg cm$^{-3}$\\
\hline \\
\end{tabular} 
\vskip 1mm
Notes: Coordinates are in B1950.0\,. Flux density and radio power are monochromatic and at 1.4 GHz. $B_{eq}$ and $u_{eq}$ are the equipartion magnetic fieldstrength and energy density, respectively, and are calculated using the assumptions made in Miley (1980).
\end{table}
  
\subsubsection{The western radio structure}

In the western radio structure, the two northern components (A and B, see Fig. \ref{fig:1144.west}) are most likely a separate radio source. The evidence for this is threefold. 
First, the radio source consisting only of components A and B put together has a morphology which resembles that of a `fat' double-lobed radio source. 
Second, the observed $E$-field polarization vectors turn around at the southern edge of component B. Such a signature is normally seen at the outer edge of FRII-type radio sources (e.g. Saikia \& Salter 1988). 
Third, on the DSS a bright ($m_R = 15.3 \pm 0.5$, estimated from the plates) galaxy is situated at a position between the radio components A and B (see Fig. \ref{fig:B1144+353.dss}). Its B1950.0 coordinates, obtained by fitting a Gaussian to the DSS image, are 11$^h$\,44$^m$\,31\fs25 in R.A. and $+35$\degr\,20\arcmin\,17\farcs5 in Dec., with an estimated uncertainty of $\sim\!3\arcsec$. 
In the FIRST survey there is no radio source at the position of the optical galaxy, which implies that the flux density of a possible radio core at 1.4 GHz is $\la 0.5$~mJy ($3\sigma$ limit). Still, on the map showing the spectral index between the WENSS and NVSS surveys (see Fig. \ref{fig:1144.wenss+nvss}a) there appears to be a region with a slightly flatter spectrum between components A and B, as compared to the spectrum of those components. This can be expected if a flat spectrum radio component, such as the radio core or a jet, is situated there. 
 We have obtained a spectrum of the optical galaxy with the 2.5-m Isaac Newton Telescope on La Palma on February 24, 1998. We used the IDS spectrograph equipped with a 1k$\times$1k TEK chip and the R158V grating which gives a dispersion of 6.56\AA~per pixel. The slitwidth was 2\arcsec, equivalent to 2.7 pixels on the CCD, and the central wavelength was 6500\AA. The total integration time was 10 min. 
The resulting spectrum is presented in Fig. \ref{fig:B1144+353.spectrum}. We have not corrected for galactic extinction. We detect H$\alpha$+$[$N{\sc ii}$]$ emission with an equivalent width of $10\pm1$\,\AA,~strongly suggesting that this galaxy indeed harbours an AGN, albeit not a powerful one. Since there is no other obvious host galaxy candidate in the DSS, it must be the host galaxy of the radio source formed by components A and B. Its redshift is not significantly different from that of the host galaxy of the GPS source ($z = 0.065 \pm 0.001$, versus $z=0.063 \pm 0.002$ for the GPS host galaxy; Marcha et al. 1996). The projected distance to the GPS host galaxy is $\sim\!380$ kpc. Therefore it probably belongs to the same group of galaxies as the GPS host galaxy. 
From the integrated flux density of components A and B we have calculated a radio power at 1.4 GHz of $1.3 \times 10^{24}$ W\,Hz$^{-1}$. The absolute magnitude of the host in the R-band is $\sim -22.8$. In a diagram of radio versus optical luminosity (see, e.g., Owen 1993), it is situated in the region occupied by low-luminosity FRI-type sources. Indeed, Owen \& White (1991) find that most sources with a `fat double' morphology lie in this region of the diagram.
Following the IAU-convention, we will refer to this radio source as B1144+353. Several of its parameters can be found in Tab. \ref{tab:B1144+353}.\\

The southern part of the western source (i.e. radio component C and the southern tail) is either associated with the host galaxy of the GPS source, or is part of the radio source B1144+353, or is a separate radio source. 
The lack of an association with an optical galaxy on the DSS rules out the last explanation. If it were associated with B1144+353, this radio source would have a highly unconventional morphology: it would be highly asymmetrical and it would have double radio lobes on one side of its nucleus. 
The bending of the polarization vectors at the southern edge of component B is characteristic of an outward moving shock which compresses the magnetic fields, and normally delimits an FRII-type radio lobe. To explain component C as a part of this radio source would require a rather exotic formation scenario. If it were the result of an earlier period of activity, we would expect to observe a similar structure northward of component A. 
Also, if the jet of B1144+353 truly reaches as far as component C we would not expect to see the bending of the polarization vectors at the edge of component B. For these reasons we do not believe that component C is physically part of the radio source B1144+353.\\ 

The most likely explanation for its existence is therefore that it is related to the GPS host galaxy. There are two further arguments to support this scenario. 
First, the WSRT observations show that the $E$-field polarization vectors turn around at the western edge of radio component C, just as they do at the southern edge of radio component B. This suggests that component C is the leading edge of a FRII-type radio lobe. As seen from the western edge of component C, the locus of the polarization vectors is situated in the direction of the GPS source, rather than in the direction of B1144+353. 
Second, the FIRST radio image of the GPS source (see Fig. \ref{fig:1144.wsrt+first}) shows two jet-like components emanating from the core. The eastern `jet' is clearly pointing towards the large eastern radio lobe, and so it is most likely that the other `jet' points towards another large radio lobe, probably at a similar distance from the nucleus. Since A and B are a separate radio source, component C is the only candidate for such a lobe.\\
  
We therefore conclude that component C forms the (former) endpoint of the western jet. FIRST has not detected a hotspot in component C, which results in an upper flux density limit ($3\sigma$) of 0.5 mJy at 1.4 GHz for such a component.\\

The line connecting the hotspot in the eastern lobe with the maximum in component C does not cross the radio core but passes $\sim\!35\arcsec$, or 57 kpc, south of it. This indicates that both jets do not follow a straight path from the core to their endpoint. Another possibility is that the host galaxy has moved towards the north while it was forming the Mpc-scale radio structure. 
A typical spectral age of a Mpc-sized radio source is $10^8$ yr (e.g. Mack et al. 1998, Schoenmakers et al. 1998a), so that a velocity of the central galaxy of $\sim\!560$ km\,s$^{-1}$ would be required to explain the possible shift. This is high, but not unconceivable.  
An argument against this scenario is that the hotspot in the eastern radio lobe, which represents the most recent endpoint of the eastern jet, is situated near the southern edge of the lobe. If the host galaxy had drifted northwards, the current endpoint of the jet would be expected near the northern edge of the lobe. Therefore we believe that a displacement of the host galaxy is not a very likely explanation for its offset from the line connecting the maxima in the outer lobes.\\  

The FIRST survey and VLBI observations (Henstock et al. 1995) show that the pc-to-kpc scale radio axis has a constant position angle of $+120^{\circ}$ (measured CCW from the North). However, the WSRT observations show that the position angles of the lines connecting the core with the maxima in the lobes have different position angles (see Fig. \ref{fig:1144.wsrt+first} and Tab. \ref{tab:B1144+352}) . On the eastern side there is a change of 8\degr~in the direction of the radio axis between kpc and Mpc-scales, on the western side the change is 18\degr. Since the kpc-scale structure is so well aligned, this indicates that the jets must be bent at a large distance (i.e. farther than 50 kpc) from the core. 
Alternatively, the radio axis may have changed direction in the course of time. However, even if the former position angle was such that the outflow on the eastern side was pointed directly at the hotspot, the western jet must still have been bent by 10\degr~to have its endpoint in the maximum of the western lobe. An interesting possibility is that the western jet has been deflected by the bowshock preceeding the expanding southern lobe of B1144+353 (component A). We have however no observational evidence for this, nor the theoretical understanding to judge the feasibility of such a scenario.\\
 
The orientation of the polarization vectors in components A, B and C is close to what one would expect for such source components. This strongly suggests that there is only marginal Faraday Rotation towards the western radio structure. If the rotation were more than $\sim\!20\degr$, we would have observed a systematic offset in the polarization angles, and so we believe the rotation must be less than this. This implies that the (absolute) Rotation Measures $|RM| \la 8$ rad\,m$^{-2}$. In case of $180\degr$ ambiguities in the polarization angle, which we cannot exclude on the basis of our data alone but which are not very likely either, $RM\!\sim\!(n\cdot68)\pm 8$ rad\,m$^{-2}$, with $n$ an integer number.

\subsubsection{The southern radio tail}
 
The existence of the southern tail of the western radio lobe is enigmatic.  
Extended radio tails have also been found in the Mpc-sized radio sources NGC\,315 and NGC\,6251 (e.g. Mack et al. 1997).
One, much favoured, scenario is that these tails are material deposited in an earlier phase of the jet-activity, with the jet pointing in a different direction.
In the case of B1144+352, there is indeed some evidence for such a change in the jet direction in the history of the radio source, as has been mentioned in the previous section. 
On the other hand, the lack of a similar tail northwards of the eastern lobe argues against a changing jet axis as the primary cause of the tail in the western lobe.\\

The tail of B1144+352 resembles the diffuse radio sources some\-times found in rich galaxy clusters (e.g. R\"{o}tt\-ge\-ring et al. 1994, Feretti \& Giovannini 1995, R\"{o}ttgering et al. 1997).  
A model for the origin of these radio halo sources is presented by Ensslin et al. (1998). They find that shocks in an intra-cluster medium, caused by the merging of two clusters, can re-accelerate particles efficiently to the energies required to emit radio synchrotron emission. The radio source would then trace the shock front, hence its elongated appearance. 
A necessary condition is that the low-energy particles that are being accelerated in the shock should be available in the intra-cluster medium. Ensslin et al. suggest that they can be the remnant of a radio source that has already faded away. 
All radio halo sources known are found in clusters with high ($10^{44} - 10^{45}$ erg\,s$^{-1}$) X-ray luminosities (e.g. R\"ottgering et al. 1997). However, the X-ray data have shown that B1144+352 is not in such an environment. At most, there is an `intra-group' medium surrounding the central radio source.
This is in accordance with the apparently low Rotation Measures towards the western radio structure, since this also implies a low (column) density of thermal electrons along the line of sight.\\

Is the southern tail perhaps the poor group equivalent of the radio halo sources in rich clusters? 
The only large radio survey of poor groups of galaxies to date is presented by Burns et al. (1987), who observed 137 poor groups with the WSRT and the VLA. This revealed only one radio source with an extended tail much similar to the tail of B1144+352. However this source (B0153+053) is most likely a head-tail radio galaxy, which Burns et al. found in large numbers in their survey. It is therefore uncertain if there is a type of radio source that exists in groups of galaxies and which is an equivalent of the extended radio halo sources.

\begin{table}
\caption{\label{tab:B1144+352}Properties of the radio lobes of B1144+352. The values for the western lobe include the southern tail. D$_{\rm max}$ is the projected distance between the position of the maximum flux density in the lobe and the GPS source. $PA$ is the position angle of the line connecting the core and the maximum in the lobes, counted CCW from the North. $S_{1400}$ and $S_{325}$ are the flux densities at 1400 and 325 MHz, respectively, and $\alpha_{325}^{1400}$ is the spectral index between these frequencies. $B_{eq}$ and $u_{eq}$ are the equipartion magnetic fieldstrength and energy density of the lobes, respectively, and are calculated using the assumptions made in Miley (1980) and the flux densities at 1400 MHz. All values have been calculated assuming that the lobes are associated to the GPS source and therefore have a similar redshift.} 
\begin{tabular}{l r l l}
\hline 
 & & \multicolumn{1}{c}{Eastern lobe} & \multicolumn{1}{c}{Western lobe} \\
D$_{\rm max}$ & $[$kpc$]$ & 631 & 521 \\
$PA$ & $[\,\degr\,]$ & $+112$ & $-78$ \\
$S_{1400}$ & $[$mJy$]$ & $100.4\pm 2.3^a$ & $48.0 \pm 1.0^b$ \\
$S_{325}$$^c$ & $[$mJy$]$ & $295 \pm 11$ & $161 \pm 35^d$ \\
$\alpha_{325}^{1400}$ &  & $-0.74 \pm 0.04$ & $-0.83 \pm 0.15$ \\
$B_{eq}$ & $[\mu$G$]$ & 1.0 & 1.0 \\
$u_{eq}$ &$[$erg\,cm$^{-3}]$ & $9.2 \times 10^{-14}$ & $8.8 \times 10^{-14}$\\
\hline \\
\end{tabular}
\vskip 1mm
Notes: a - measured from the NVSS map; b - measured from the WSRT map; c - measured from the WENSS map; d - somewhat confused with the source B1144+353.
\end{table}  

\subsection{The orientation of the radio source}
\label{sec:orientation}

Giovaninni et al. (1995) have found superluminal motion in the pc-scale radio structure, with an apparent velocity of 2.4$c\,h_{50}^{-1}$. In the cosmology we assume, this implies that the angle $\theta$ between the radio axis and the line of sight must be $\leq 45.2\degr$. Considering the already large linear size of the outer structure, the lack of broad emission lines and non-thermal continuum in the optical spectrum, and the relatively low X-ray luminosity of the central source when compared with broad-line radio galaxies (e.g. Crawford \& Fabian 1995), we believe that a value close to the upper limit of $45\degr$ is more appropriate. \\

If the kpc-scale structure has a similar orientation to the pc-scale structure and if it originated in the core, the armlength asymmetry of the inner structure observed in FIRST can be explained as a pure orientation effect. We further assume that the advance velocity of the outer components is equal on both sides of the nucleus. Since the armlength ratio of the outer components in the FIRST map is $1.684$, their advance velocity must be $\sim\!0.36c$, which is much higher than what is normally found in radio galaxies ($0.01-0.1c$, e.g. Alexander \& Leahy 1987). This would suggest that the inner lobes are expanding an in extremely low-density environment, much lower than normally found at such distances from the radio core.\\

Also the flux density ratio of the kpc-scale components on either side of the core, if they are indeed advancing at this velocity, can be predicted under the hypothesis that the two sides are intrinsically equal and that any side-to-side difference is only due to relativistic beaming. 
Using the formula given in Pearson \& Zensus (1987), and assuming that the components are discrete radio sources advancing at an angle $\theta\sim\!45\degr$ with respect to the line of sight, we find an expected flux density ratio of 7.3\,. The observed flux density ratio is $4.5 \pm 0.9$\,, which, considering the simplifying assumptions, is not far off.
However, velocities of almost $0.4c$ at distances of tens of kpc from the nucleus are not expected in radio sources, and so we believe that the armlength and flux density asymmetries are caused by some other effect, most likely inhomogenities in the environment.\\
 
An alternative hypothesis to explain the large asymmetry in armlength of the inner kpc-scale structure is that of a jet which is ejected alternatively on one side of the nucleus and then on the other (e.g. Rudnick \& Edgar 1984). The observation that the radio structure on the eastern side appears to consist of two components, one at 21.6 kpc from the nucleus and the other at 41.2 kpc, while the radio structure on the western side only consists of a single component at 24.5 kpc from the core, appears to support such a scenario. This suggests that the outflow direction of the jet responsible for the formation of the inner structure would have `flipped' at least twice. 
A consequence of this model is that only one side of the radio source is fuelled by the jet. Since radio hotspots are expected to fade very rapidly once the inflow of jet material has stopped (within a few $10^4$ yr; e.g. Clarke \& Burns 1991), a large asymmetry in flux density is a natural result. 
Although statistical arguments argue against this scenario (e.g. Ensman \& Ulvestad 1984), we can not rule out the possibility that B1144+352 is a special case.\\

The projected distance from the hotspot in the eastern lobe to the maximum in component C of the western structure is $\sim\!1150$ kpc. If the Mpc-scale structure also has a similarly oriented radio axis as the pc-scale structure, the true physical size of B1144+352 is $\ga\!1.6$ Mpc, which is not an uncommon value for Giant radio sources (e.g. Saripalli et al. 1986).

\subsection{Is the nucleus recurrently radio active?}
\label{sec:recurrency}
B1144+352 is one of two currently known GPS sources which are also associated with Mpc-sized radio sources. The other case is the source B1245+676 which has outer lobes of 1.9 Mpc projected linear size (de Bruyn et al. in preparation). These two sources can be seen as the extremes of the few known GPS sources with extended radio structures, such as those discussed by Baum et al. (1990) and Stanghellini et al. (1990). 
In the case of B1144+352, the presence of sharply bound radio structures on Mpc, kpc and pc-scale suggests that the AGN must have gone through several phases of radio activity or, alternatively, that the jets of a continuously operating AGN must have been disrupted (or `smothered'; Baum et al. 1990) several times close to the nucleus.
The fact that the core is currently a GPS source, and so either must be very young, or entrapped in a high density environment, adds further weight to the argument that some form of interruption of the jet flow must have occurred. 
Since superluminal motion has been detected on pc-scales, it is unlikely that the GPS source is confined by a high density medium. Therefore, the most likely scenario for this source is that the nucleus is indeed recurrently radio active.  O'Dea (1998) mentions that the large fraction of GPS/CSS galaxies that show evidence for interaction and/or mergers, suggests that such processes must be related with the formation of the radio sources. An apparent close companion is also detected near the host galaxy of B1144+352 (see Sect. \ref{sec:1144.intro} and Fig. \ref{fig:1144_dss}). Perhaps an interaction between the host galaxy and this apparent companion has disturbed the jet flow to the outer lobes, or even halted the jet formation temporarily, giving rise to the subsequent formation of the inner radio structure.\\

The rise and fall in the 1.4-GHz flux density of the central source over the last 25 years and the presence of the luminous X-ray source suggest that the nucleus is currently active. 
The distance between the outermost maximum in the eastern inner kpc-structure and the core is $41.2/\sin\theta$ kpc (see Tab. \ref{tab:1144_first}), where  $\theta$ is the angle between the radio axis and the line of sight. This indicates that the nucleus must have been radio active, although not necessarily continuously, for at least the last $1.3\times10^5 c/(v_{adv} \sin\theta$) yr. Here, $v_{adv}$ is the advance velocity of the head of the jet.
The presence of the GPS source implies that jet material which is currently expelled from the core does not flow into the outer Mpc-sized lobes anymore, and possible also not into the kpc-sized lobes either. Still, a compact radio hotspot has been detected in the eastern Mpc-sized radio lobe of B1144+352. 
As has been mentioned in the last section, radio hotspots are expected to fade within a few $10^4$ yr (Clarke \& Burns 1991). Therefore, the detection of the hotspot implies that there must still be jet material arriving there. 
This means that the disruption of the jet cannot have occurred longer ago than the travel-time of the jet material from the nucleus to this hotspot. The travel-time is $2\times10^6\,c/(v_j\,\sin \theta)$ yr, where $v_j$ is the velocity of the material flowing down the jet and which is most likely close to $c$. This also gives an maximum age of the inner structure.
Given the constraint on the age, we find that the minimal life-time averaged advance velocity of the head of the inner structure, $v_{adv}$, is $0.07c$. Depending on how long the radio activity was switched off this value further increases.\\

If, eventually, the central kpc-scale radio source grows out to a large ($\sim\!100$ kpc) radio source before the outer lobes have faded, B1144+352 can become a so-called `double-double radio galaxy' (DDRG, Schoenmakers et al. 1998b). DDRGs are radio sources which consist of two unequally sized but well aligned double-lobed radio sources with a coinciding radio core. Six of such objects with inner sources larger than 100~kpc are known now (Schoenmakers et al. 1998b), and they also appear to be the result of recurrent jet-formation activity in the nucleus. 
Sources such as B1144+352 and B1245+676 are good candidates for the progenitors of these DDRGs.

\section{Conclusions}

We have studied the large-scale radio structure of nearby GPS source B1144+352. We have presented new 1.4-GHz WSRT observations, and we have obtained X-ray data from the ROSAT archive. Our conclusions are the following:\\
\begin{enumerate}
\item After a steady increase over at least 25 yr, the core flux density at 1.4 GHz has decreased by $\sim\,70$ mJy to $541 \pm 10$ mJy (epoch 1997.7) with respect to the measurements in the FIRST survey (epoch 1994.6). This behaviour is remarkable since GPS galaxies in general show only small variations in their flux density, but agrees with the predictions made by a model of the evolutionary scenario of radio components in GPS sources presented by Snellen et al. (1998). 
\item The eastern extended radio structure is a radio lobe with a leading hotspot and can only be associated with the central GPS source.
\item The southern part of the western structure is a radio lobe with an elongated tail. This lobe is most likely associated with the GPS source. The origin of the southern tail is not clear.
\item The total projected linear size of the large-scale radio source associated with the GPS source B1144+352 is $\sim\!1.2$ Mpc, but the detection of superluminal motion in the GPS source suggests that the true (deprojected) physical size may be $\ga\!1.6$ Mpc.
\item There are differences of $\sim\!10\degr$ between the pc-to-kpc-scale radio axis and the Mpc-scale radio axis. This means that the Mpc-scale jets must have been bent. It also suggests that the outflow direction of the jet responsible for the pc-to-kpc-scale structure has changed before the formation of the kpc-scale structure.
\item  Since we still see a compact hotspot in the eastern Mpc-scale radio lobe, we can limit the age of the kpc-scale radio source to $\la 2\times10^6 / \sin\theta$ yr, with $\theta$ the angle of the radio axis with respect to the line of sight. Therefore, the time-averaged advance velocity of the eastern kpc-scale component must be $\ga 0.07c$. 
\item The host galaxy of B1144+352 is associated with a somewhat extended X-ray source. The extension appears to have a similar position angle to the central radio source. Assuming that the emission is due to an AGN, with a powerlaw spectrum with photon index 1.8, we find a luminosity of $(1.28\pm0.15)\times10^{43}$ erg\,s$^{-1}$ between 0.1 and 2.4 keV. For a thermal bremsstrahlung model with $T = 10^7$ K and a metal abundance of 0.25 solar, we find a luminosity of $(0.95\pm0.11) \times10^{43}$ erg\,s$^{-1}$ in the same band. Such luminosities are comparable to that of powerful 3CR narrow-line radio galaxies at similar redshift. It strongly suggests that the central part of the host galaxy of B1144+352 does not contain a large amount of cold gas.
\item In the environment of the GPS source some sort of medium must be present to confine the extended radio structures. Such a medium may reveal itself through the emission of X-rays. Using the lack of a significant detection in our ROSAT data, we have determined a 1$\sigma$ upper limit of $3.2 \times 10^{42}$ erg\,s$^{-1}$ for the $(0.5-2)$ keV X-ray luminosity within a 250 kpc radius of the host galaxy of B1144+352. Therefore, the radio source is not in a luminous X-ray cluster. The upper limit of the X-ray luminosity is, however, comparable to that of poor groups of galaxies, so we cannot exclude the presence of an `intra-group' medium. Optical data seem to support the presence of a small group of galaxies around the host galaxy of B1144+352.
\item The northern part of the western extended radio structure is a separate low-power radio galaxy which we refer to as B1144+353. We have measured the redshift of the host galaxy at $0.065 \pm 0.001$, which is not significantly different from that of the GPS host galaxy ($0.063 \pm 0.002$). The projected distance to the GPS host galaxy is $\sim\!380$ kpc.\\ 
\end{enumerate}

The source B1144+352 is the second GPS source known to be associated with a Mpc-sized radio source. Notwithstanding whether GPS sources are young or situated in dense environments, the existence of the Mpc-sized lobes implies that the jets fuelling the radio source have been able to escape the host galaxy in the past. 
We conclude that the best explanation for the existence of these two sources is that their jet flow must have been interrupted temporarily, either as a result of a complete halting of the jet formation or as a result of a disruption of the jet flow close to the nucleus. This makes it very likely that they are the progenitors of the DDRGs discussed by Schoenmakers et al. (1998b). 
In case of the source B1144+352, the observation of superluminal motion on pc-scale indicates that the GPS source it is not confined, or `frustrated', by a high density medium, and that therefore the nucleus is most likely recurrently radio active. The variation in radio peak flux density observed over the last 25 years suggests that a new phase of activity may have started quite recently and that the GPS source which is currently observed is extremely young.

\begin{acknowledgements}
The authors would like to thank I. Snellen, for pointing out some of the interesting properties of this GPS source, and F. Verbunt, for his help with the analysis of the ROSAT data. We would like to thank the referee, A. Cooray, for extensive comments that improved the paper considerably.
The INT is operated on the island of La Palma by the Isaac Newton Group in the Spanish Observatorio del Roque de los Muchachos of the Instituto de Astrofisica de Canarias. We thank G. Cotter for his assistance during the observations.
The Westerbork Synthesis Radio Telescope (WSRT) is operated by the
Netherlands Foundation for Research in Astronomy (NFRA) with financial
support of the Netherlands Organization for Scientific Research (NWO).
The National Radio Astronomy Observatory (NRAO) is operated by Associated Universities, Inc., and is a facility of the National Science Foundation (NSF).
This research has made use of the NASA/IPAC Extragalactic Database (NED) which is operated by the Jet Propulsion Laboratory, California Institute of Technology, under contract with the National Aeronautics and Space Administration.
The Digitized Sky Surveys were produced at the Space Telescope Science Institute under U.S. Government grant NAG W-2166. The images of these surveys are based on photographic data obtained using the Oschin Schmidt Telescope on Palomar Mountain and the UK Schmidt Telescope. The plates were processed into the present compressed digital form with the permission of these institutions. 

\end{acknowledgements}

{}


\begin{thebibliography}{}

\bibitem{} Alexander P., Leahy J.P., 1987, MNRAS 225, 1
\bibitem{} Baars J.W.M., Genzel R., Pauliny-Toth I.I.K., Witzel A., 1977, A\&A 61, 99
\bibitem{} Baum S.A., O'Dea C.P., Murphy D.W., de Bruyn A.G., 1990, A\&A 232, 19
\bibitem{} Becker R., White R., Helfand D., 1995, ApJ 450, 559
\bibitem{} Begelman M.C., 1998, In: The High-redshift Universe, Best P., Lehnert M., R\"{o}ttgering H. (eds.), Kluwer (Dordrecht), in press
\bibitem{} Bicknell G.V., Dopita M.A., O'Dea C.P., 1997, ApJ 485, 112
\bibitem{} van Breugel W.J.M., Miley G., Heckman T., 1984, AJ 89, 5
\bibitem{} Burns J.O., Hanisch R.J., White R.A., Nelson E.R., Morrisette K.A., Moody J.W., 1987, AJ 94, 587
\bibitem{} Carilli C.L., Menten K.M., Reid M.J., Rupen R.P., Yun M.S., 1998, ApJ 494, 175
\bibitem{} Clarke D.A., Burns J.O., 1991, ApJ 369, 308   
\bibitem{} Colla G., Fanti C., Fanti R., Gioia I., Lari C., Lequeux J., Lucas R., Ulrich M., 1975a, A\&AS 20, 1
\bibitem{} Colla G., Fanti C., Fanti R., Gioia I., Lari C., Lequeux J., Lucas R., Ulrich M., 1975b, A\&A 38, 209
\bibitem{} Condon J.J., Cotton W.D., Greisen E.W., Yin Q.F., Perley R.A., Taylor G.B., Broderick J.J., 1998, AJ 115, 1693
\bibitem{} Crawford C.S., Fabian A.C., 1995, MNRAS 273, 827
\bibitem{} Dallacasa D., Fanti C., Fanti R., Schilizzi R.T., Spencer R.E., 1995, A\&A 295, 27
\bibitem{} David L.P., Harnden F.R., Kearns K.E., Zombeck M.V., 1993, The ROSAT High Resolution Imager, US ROSAT Science Data Center, SAO
\bibitem{} Ensslin T.A., Biermann P.L., Klein U., Kohle S., 1998, A\&A 332, 395
\bibitem{} Ensman L.M., Ulvestad J.S., 1984, AJ 89, 1275
\bibitem{} Fanaroff B.L., Riley J.M., 1974, MNRAS 167, 31
\bibitem{} Fanti C., Fanti R., de Ruiter H.R., Parma P., 1987, A\&AS 65, 145
\bibitem{} Fanti C., Fanti R., Dallacasa D., Schilizzi R.T., Spencer R.E., Stanghellini C., 1995, A\&A 302, 117
\bibitem{} Feretti L., Giovannini G., 1995, In: IAU Symposium 175: Extragalactic Radio Sources, Eds. Ekers R., Fanti C., Padrielli L., Kluwer Academic Publishers, p. 333
\bibitem{} Forman W., Jones C., 1991. In: Clusters and superclusters of galaxies, Ed. Fabian A., Kluwer Academic Press, Dordrecht, p. 49
\bibitem{} Giovannini G., Feretti L., Comoretto G., 1990, ApJ 358, 159 
\bibitem{} Giovannini G., Cotton W.D., Feretti L., Lara L., Venturi T., Marcaide J.M., 1995, In: Proceedings of N.A.S. Colloquium "Quasars and AGN: High resolution Radio Imaging", Irvine, eds. Cohen M.H. and Kellermann K.I.
\bibitem{} Hartmann D., 1994, Ph.D. thesis, University of Leiden
\bibitem{} Henstock D.R., Browne I.W.A., Wilkinson P.N., Taylor G.B., Vermeulen R.C., Pearson T.J., Readhead A.C.S., 1995, ApJS 100, 1
\bibitem{} Hewitt A., Burbidge G., 1991, ApJS 75, 297
\bibitem{} Jones L.R., Scharf C., Ebeling H., Perlman E., Wegner G., Malkan M., Horner D., 1998, ApJ 495, 100
\bibitem{} Mack K.-H., Klein U., O'Dea C.P., Willis A.G., 1997, A\&AS 123, 423
\bibitem{} Mack K.-H., Klein U., O'Dea C.P., Willis A.G., Saripalli L., 1998, A\&A 329, 431
\bibitem{} Marcha M.J.M., Browne I.W.A., Impey C.D., Smith P.S., 1996, MNRAS 281, 425 
\bibitem{} Miley G.K., 1980, ARA\&A 18, 165
\bibitem{} Mukai K., 1993, Legacy 3, 21
\bibitem{} Mulchaey J.S., Zabludoff A.I., 1998, ApJ 496, 73
\bibitem{} Mulchaey J.S., Davis D.S., Mushotzky R.F., Burstein D., 1996, ApJ 456, 80
\bibitem{} O'Dea C.P., 1998, PASP 110, 493
\bibitem{} O'Dea C.P., Baum S.A., 1997, AJ 113, 148
\bibitem{} O'Dea C.P., Baum S.A., Stanghellini C., 1991, ApJ 380, 66
\bibitem{} O'Dea C.P., Stanghellini C., Baum S.A., Charlot S., 1996a, ApJ 470, 806
\bibitem{} O'Dea C.P., Worall D.M., Baum S.A., Stanghellini C., 1996b, AJ 111,92
\bibitem{} Owen F.N., 1993, In: Jets in extragalactic radio sources, eds. Rosen H.-J., Meisenheimer K., Springer New-York, p. 273--278
\bibitem{} Owen F.N., White R.A, 1991, MNRAS 249, 164
\bibitem{} Owsianik I., Conway J.E., Polatidis A.G., 1998, A\&A 336, L37
\bibitem{} Pacholczyk A.G., 1970, Radio Astrophysics, Freeman, San Francisco
\bibitem{} Parma, P., de Ruiter H.R., Fanti C., Fanti R., 1986, A\&AS 64, 135
\bibitem{} Pearson T.J., Zensus J.A., 1987, In: Superluminal Radio Sources, eds. Pearson T.J., Zensus J.A., Cambridge University Press, p. 1--11  
\bibitem{} Phillips R.B., Mutel R.L., 1982, A\&A 106, 21
\bibitem{} Readhead A.C.S., Taylor G.B., Xu W., Pearson T.J., Wilkinson P.N., Polatidis A.G., 1996, ApJ 460, 612 
\bibitem{} Rengelink R., Tang Y., de Bruyn A.G., Miley G.K., Bremer M.N., R\"{o}ttgering H.J.A., Bremer M.A.R., 1997, A\&AS 124,259 
\bibitem{} Reynolds C.S., Begelman M.C., 1997, ApJ 487, L135 
\bibitem{} R\"{o}ttgering H.J.A., Snellen I.A.G., Miley G.K., de Jong J., Hanisch B., Perley R., 1994, ApJ 436, 654
\bibitem{} R\"{o}ttgering H.J.A., Wieringa M.H., Hunstead R.W., Ekers R.D., 1997, MNRAS 290, 577
\bibitem{} Rudnick L., Edgar B.K., 1984, ApJ 279, 74
\bibitem{} Saikia D.J., Salter C.J., 1988, ARA\&A 26, 93
\bibitem{} Saripalli L., Gopal-Krishna, Reich W., K\"{u}hr H., 1986, A\&A, 170, 20
\bibitem{} Scott M.A., Readhead A.C.S., 1977, MNRAS 180, 593
\bibitem{} Schoenmakers A.P., Mack K.-H., Lara L., R\"{o}ttgering H.J.A., de Bruyn A.G., van der Laan H., Giovannini G., 1998a, A\&A 336, 455
\bibitem{} Schoenmakers A.P., R\"{o}ttgering H.J.A., de Bruyn A.G., van der Laan, H., 1998b, {\sl submitted to MNRAS}
\bibitem{} Snellen I.A.G., 1997, PhD-thesis, University of Leiden
\bibitem{} Snellen I.A.G., Zhang M., Schilizzi R.T., R\"{o}ttgering H.J.A., de Bruyn A.G., Miley G.K., 1995, A\&A 300, 359 (S95)
\bibitem{} Snellen I.A.G., Bremer M.N., Schilizzi R.T., Miley G.K., van Ojik R., 1996, MNRAS 279, 1294 
\bibitem{} Snellen I.A.G., Schilizzi R.T., de Bruyn A.G., Miley G.K., 1998, A\&A 333, 70 
\bibitem{} Spoelstra T.A.T., Patnaik A.R., Gopal-Krishna, 1985, A\&A 152, 38
\bibitem{} Stanghellini C., Baum S.A., O'Dea C.P., Morris G.B., 1990, A\&A 233, 379
\bibitem{} Stanghellini C., O'Dea C.P., Baum S.A., Dallacasa D., Fanti R., Fanti C., 1997, A\&A 325, 943
\bibitem{} Stanghellini C., O'Dea C.P., Baum S.A., Dallacasa D., Fanti R., Fanti C., 1998, A\&AS 131, 303
\bibitem{} White R.L., Becker R.H., 1992, ApJS 79, 331
\bibitem{} Williams O.R., et al., 1992, ApJ 389, 157
\bibitem{} Wilkinson P.N., Polatidis A.G., Readhead A.C.S., Xu W., Pearson T.J., 1994, ApJ 432, L87
\bibitem{} Worrall D.M., Birkinshaw M., 1994, ApJ 427, 134.  
\bibitem{} Zwicky F., Herzog E., Karpowicz M., Kowal C.T., Wilp P., 1960-1968, Catalogue of galaxies and clusters of galaxies, California Institute of Technology, Vols. 1--6 (CGCG)
\end{thebibliography}
\end{document}